\documentclass[runningheads]{llncs}
\usepackage[T1]{fontenc}
\usepackage{graphicx}
\usepackage{amsmath}
\usepackage{amssymb}
\usepackage{xspace}
\usepackage{tikz}
\usepackage{graphicx}
\usepackage{multirow}
\usepackage{pdfrender}
\usepackage{color, colortbl}

\definecolor{LightCyan}{rgb}{0.88,1,1}
\usepackage[pagebackref,breaklinks,colorlinks,citecolor=blue]{hyperref}
\usepackage[capitalize]{cleveref}
\crefname{section}{Sec.}{Secs.}
\Crefname{section}{Section}{Sections}
\Crefname{table}{Table}{Tables}
\crefname{table}{Table}{Tables}
\def\R{\mathbb{R}}

\newcommand{\bd}[1]{\boldsymbol{#1}}
\newcommand{\best}[1]{\textcolor{red}{\underline{#1}}}
\newcommand{\sbest}[1]{\textcolor{blue}{\textit{#1}}}
\makeatletter
\DeclareRobustCommand\onedot{\futurelet\@let@token\@onedot}
\def\@onedot{\ifx\@let@token.\else.\null\fi\xspace}
\newcommand*{\boldcheckmark}{%
  \textpdfrender{
    TextRenderingMode=FillStroke,
    LineWidth=.5pt, 
  }{\checkmark}%
}

\def\ie{\emph{i.e}\onedot} 
 
\def\etc{\emph{etc}\onedot}

\makeatother
\begin{document}
\title{MoME: Mixture of Multimodal Experts for Cancer Survival Prediction}
%
\titlerunning{MoME: Mixture of Multimodal Experts for Cancer Survival Prediction}

\author{Conghao Xiong\inst{1}\thanks{Corresponding authors.} 
\and Hao Chen\inst{2} 
\and Hao Zheng\inst{3}$^\star$ 
\and Dong Wei\inst{3} 
\and Yefeng Zheng\inst{3} 
\and Joseph~J.~Y.~Sung\inst{4} 
\and Irwin King\inst{1}} 
\authorrunning{C. Xiong et al.}
%
\institute{Dept.~of Computer Science and Engineering, The Chinese University of Hong Kong \and
Dept.~of Computer Science and Engineering and Dept.~of Chemical and Biological Engineering, The Hong Kong University of Science and Technology \and 
Jarvis Research Center, Tencent YouTu Lab \and
Lee Kong Chian School of Medicine, Nanyang Technological University
}
\maketitle              
\begin{abstract}
Survival analysis, as a challenging task, requires integrating Whole Slide Images (WSIs) and genomic data for comprehensive decision-making. There are two main challenges in this task: significant heterogeneity and complex inter- and intra-modal interactions between the two modalities. Previous approaches utilize co-attention methods, which fuse features from both modalities only once after separate encoding. However, these approaches are insufficient for modeling the complex task due to the heterogeneous nature between the modalities. To address these issues, we propose a Biased Progressive Encoding (BPE) paradigm, performing encoding and fusion simultaneously. This paradigm uses one modality as a reference when encoding the other. It enables deep fusion of the modalities through multiple alternating iterations, progressively reducing the cross-modal disparities and facilitating complementary interactions. Besides modality heterogeneity, survival analysis involves various biomarkers from WSIs, genomics, and their combinations. The critical biomarkers may exist in different modalities under individual variations, necessitating flexible adaptation of the models to specific scenarios. Therefore, we further propose a Mixture of Multimodal Experts (MoME) layer to dynamically selects tailored experts in each stage of the BPE paradigm. Experts incorporate reference information from another modality to varying degrees, enabling a balanced or biased focus on different modalities during the encoding process. Extensive experimental results demonstrate the superior performance of our method on various datasets, including TCGA-BLCA, TCGA-UCEC and TCGA-LUAD. Codes are available at \url{https://github.com/BearCleverProud/MoME}.

\end{abstract}
\section{Introduction}
Survival analysis via \textbf{W}hole \textbf{S}lide \textbf{I}mages (WSIs) and genomic data is crucial in pan-cancer prognosis as it assesses the risk of death and provides important references for the treatment plans. The key to this task is how to effectively utilize information from both modalities, for instance, to detect image-omic biomarkers as well as to explore interactions between tumor microenvironment in histopathology images and co-expression of genomic data. In recent years, the focus of research also has shifted from single-modal prediction \cite{campanella_clinical-grade_2019,ianni_tailored_2020,klambauer_self_2017,litjens_deep_2016,xiong2023knowledge} to the more complicated survival analysis utilizing multimodal information \cite{chen2021multimodal,ding2023pathology,xing2022discrepancy,xu_2023_motcat,zhou2023cross}.

One of the key challenges in this task is the significant heterogeneity between histopathology images and genomic data \cite{li_hfbsurv_2022}, stemming from their inherent disparities and distinct pre-processing methods. Additionally, the inter- and intra-modal interactions are highly complex, as both modalities possess abundant information, but only a small fraction of them can be mutually correlated and utilized for survival prediction. Previous approaches have attempted to tackle this challenge by using cross-modality attention (co-attention) \cite{xu_2023_multimodal} based methods \cite{chen2021multimodal,xu_2023_motcat,zhou2023cross}. However, feature fusion is conducted only once throughout the entire process. These approaches might be considered shallow given the complexity of the task and the significant differences between the two modalities.

To tackle these issues, we propose a \textbf{B}iased \textbf{P}rogressive \textbf{E}ncoding (BPE) paradigm. Unlike previous methods that encode modalities separately before fusion, our approach simultaneously encodes and fuses features. In this approach, one modality is encoded while utilizing the other modality as a reference, which can assist in extracting more relevant information. Furthermore, the encoding of features from the two modalities is performed alternately, thereby progressively reducing the differences between their feature spaces. This allows for a deeper fusion process and facilitates the discovery of interactions between modalities. 

In addition to modality heterogeneity, inter-individual variations can cause key features for survival analysis to appear in different modalities for each patient. This presents a new challenge in designing the model structure, as it requires selective focus on a specific modality or the interactions between the two modalities. To achieve this, we propose a \textbf{M}ixture \textbf{o}f \textbf{M}ultimodal \textbf{E}xperts (MoME) layer, employing our BPE paradigm. The MoME layer consists of multiple specialized experts capable of modeling complex inter- and intra-modal correlations. In addition, these experts incorporate reference information from another modality to varying degrees, enabling a focus on different modalities during the encoding process. Moreover, we enable flexible selection of experts in each layer, as the function of reference information may differ. In the shallow layers of the network, our MoME layer could use the reference information as a filter to eliminate task-irrelevant features and enhance relevant ones within each modality. Conversely, in the deeper layers of the network, it could be used as a guidance to seeks cross-modal combination representations as biomarkers.

The contributions of this paper can be summarized as follows:
\begin{enumerate}
    \item We propose a biased progressive encoding paradigm which integrates information from one modality into the feature encoding of the other modality as a reference for more effective feature extraction and interaction modeling.
    \item We design a mixture of multimodal experts layer which enables the network to selectively focus on the information from a specific modality and utilizes the reference information in different forms across encoding stages.
    \item We extensively evaluate our method on three TCGA datasets: BLCA, UCEC, and LUAD. The results demonstrate the superior performance of our method.
\end{enumerate}

\section{Methodology}
\subsection{Problem Formulation}
The WSIs are usually formulated under the \textbf{A}ttention-\textbf{B}ased \textbf{M}ultiple \textbf{I}nstance \textbf{L}earning (AB-MIL) framework \cite{ilse_attention-based_2018,xiong_diagnose_2023}. This involves dividing the WSIs into bags of patches and extracting features from them using a pre-trained neural network. An MIL aggregator is then used to process the features, generate bag features, and make predictions. The WSI patch features are represented as $\bd{P} \in \R^{n_p\times d}$, where $n_p$ is the number of patches in the WSI bag and $d$ is the dimension for both WSI and genomic latent features.
The genomic data consists of various $1\times 1$ values, including RNA sequencing, copy number variation, DNA methylation, \etc. Following previous works \cite{chen2021multimodal}, the genomic data are categorized into the following genomic sequences: 1) Tumor Suppression, 2) Oncogenesis, 3) Protein Kinases, 4) Cellular Differentiation, 5) Transcription, and 6) Cytokines and Growth. These sequences are stacked and fed into a fully connected layer to obtain genomic features, which are denoted as $\bd{G} \in \R^{n_g \times d}$, where $n_g=6$ is the number of genomic sequences. 
In survival analysis, given the input pair ($\bd{P}$,$\bd{G}$), rather than predicting the exact time of death for patients, we initially estimate the hazard function $h(t) = h(T=t|T\ge t, (\bd{P},\bd{G})) \in [0,1]$, which is the probability of death for a patient right after the time point $t$. Subsequently, an ordinal value is obtained via integrating the negated hazard functions: $s(t|(\bd{P},\bd{G})) = \prod_{u=1}^t(1-h(u))$. 
\begin{figure}[t]
    \centering
    \includegraphics[width=\linewidth]{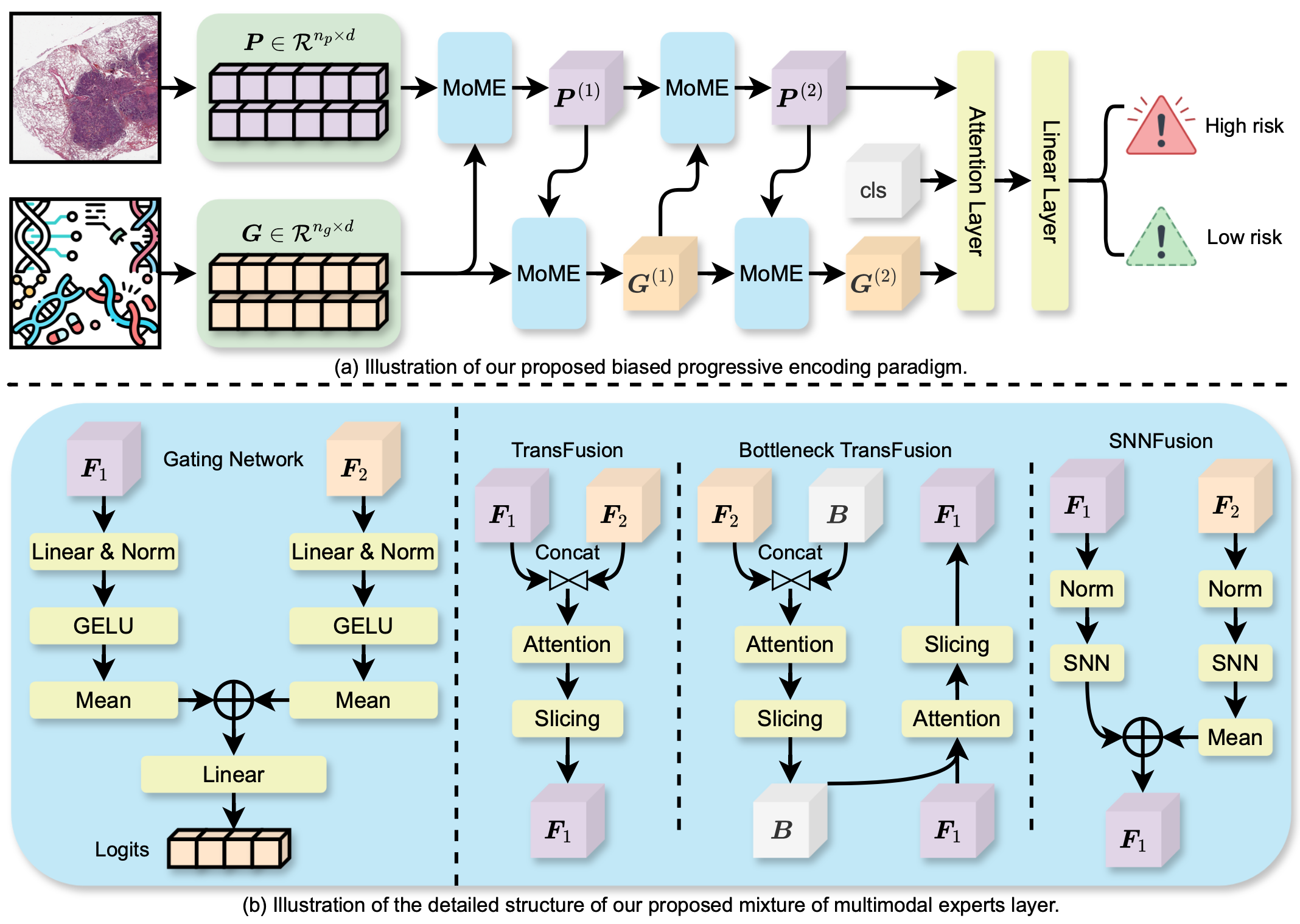}
    \caption{Illustrations of (a) our biased progressive encoding paradigm, and (b) the structure of our mixture of multimodal experts. The left section of (b) represents our gating network, and the right section of (b) depicts our proposed three expert components designed for different degrees of integration of the reference modality.}
    \label{fig:framework}
\end{figure}
\subsection{Biased Progressive Encoding Paradigm}
The overview of our BPE paradigm is shown in \cref{fig:framework}(a). For simplicity, we refer to the modality being encoded as $\bd{F}_1^{(i)}$, and the reference modality as $\bd{F}_2^{(i)}$, where $\bd{F}_1$ could be either $\bd{P}$ or $\bd{G}$, and $i$ is the times of encoding. Our BPE paradigm follows a progressive learning strategy, where $\bd{F}_1^{(i)}$ is encoded to discover the complex interactions with $\bd{F}_2^{(i)}$ being a reference. This process is then reversed to encode the other modality $\bd{F}_2^{(i)}$ with the encoded first modality $\bd{F}_1^{(i+1)}$. The deep feature extraction and progressive learning strategy in our BPE enable deep fusion to reduce the inter-modal discrepancy. The complete progressive encoding process, which involves encoding both modalities, is given as,
\begin{equation}
        \bd{F}_1^{(i+1)} = \operatorname{BPE}_{2i}(\bd{F}_1^{(i)}, \bd{F}_2^{(i)}), \quad \quad\bd{F}_2^{(i+1)} = \operatorname{BPE}_{2i+1}(\bd{F}_2^{(i)}, \bd{F}_1^{(i+1)}),
\end{equation}
where $\operatorname{BPE}_{\cdot}(\cdot, \cdot)$ is our MoME, detailed in the next section. We encode both features twice for all datasets, facilitating a deep fusion of the two modalities. Following the encoding, both features are fed into an attention block along with a classification token, which is then used for the final survival prediction.

\subsection{Mixture of Multimodal Experts}
The structure of our MoME layer is depicted in \cref{fig:framework}(b). Our MoME layer is derived from the traditional \textbf{M}ixture \textbf{o}f \textbf{E}xperts (MoE) \cite{william2022switch,masoudnia_2014_moe_survey,noam2017outrageously}, consisting of a set of parallel feed forward networks (experts) and a gating network that controls the selection of experts. Distinct from the classic MoE that operate at the token level, routing tokens within a sequence to various experts, our MoME innovates by functioning at the sample and layer levels, \ie, different samples within the same layer or identical samples across different layers can be routed to distinct experts. This innovation is pivotal for handling the informative yet sparse characteristics of WSIs and genomics, where isolated features may be nondescript, underscoring the necessity to process them collectively.

Our MoME comprises two components: 1) a gating network and 2) a set of specially designed experts for multimodal survival analysis. Initially, the features are passed to the gating network, which determines the most suitable expert to utilize. The selected expert then performs fusion and encoding for $\bd{F}_1^{(i)}$.

\subsubsection{Gating Network.} 
The gating network is designed to be lightweight yet informative to select experts. It consists of linear layers, \textbf{G}aussian \textbf{E}rror \textbf{L}inear \textbf{U}nits (GELUs) \cite{hendrycks2016gaussian}, and \textbf{R}ooted \textbf{M}ean \textbf{S}quare Layer \textbf{Norm}alization (RMSNorm) layers \cite{zhang_2019_rmsnorm}. These modules map features into a new space, and the mapped features are averaged to obtain the multimodal one. The gating network is given as,
\begin{equation}
    \begin{aligned}
        \operatorname{logits} =\ &\bd{W}\cdot(\operatorname{mean}(\operatorname{GELU}(\operatorname{RMSNorm}(\bd{W}_1\bd{F}_1)))) \\
        &+ \bd{W}\cdot(\operatorname{mean}(\operatorname{GELU}(\operatorname{RMSNorm}(\bd{W}_2\bd{F}_2)))),
    \end{aligned}
\end{equation}
where $\bd{W}_1, \bd{W}_2$ and $\bd{W}$ are learnable matrices. The obtained logits are utilized to select the appropriate expert. Unlike traditional MoE models that employ a weighted sum of multiple experts \cite{masoudnia_2014_moe_survey}, our approach enforces the selection of only one expert within the module \cite{xue_dynamic_2023}. By adopting this strategy, the gating network can make careful expert choices while also reducing computational costs.

\subsubsection{Multimodal Expert Pool.} 
We design the following four experts based on the two principles: 1) inclusion of experts specializing in WSI, genomics, and interactions between them, and 2) capability of simultaneous fusion and encoding. 

\paragraph{TransFusion.} This multimodal expert is based on self-attention. This expert maximizes the utilization of the reference modality by enabling full information exchange between the two modalities with self-attention. Given the input pair $(\bd{F}_1,\bd{F}_2)$, our proposed \textbf{T}rans\textbf{F}usion (TF) expert can be expressed as, 
\begin{equation}
    \operatorname{TF}(\bd{F}_1, \bd{F}_2) = \operatorname{SA}([\bd{F}_1, \bd{F}_2])[:n_1,:],
\end{equation}
where $\operatorname{SA}(\cdot)$ is the \textbf{S}elf-\textbf{A}ttention \cite{transformer_2017}, $[\bd{F}_1, \bd{F}_2] \in R^{(n_1+n_2)\times d}$ is the concatenation of $\bd{F}_1$ and $\bd{F}_2$, $[:n_1,:]$ indicates to select the first $n_1$ rows of the matrix, and $n_1,n_2$ denote the number of features of the two modalities $\bd{F}_1$ and $\bd{F}_2$.

\paragraph{Bottleneck TransFusion.} Both modalities contain a vast amount of information, but only a small portion of them is useful for survival prediction. Therefore, it is necessary to distill the information and focus only on the portions that are pertinent to the survival analysis outcome. To tackle this, we propose an expert that avoids direct interactions between the two modalities. Instead, we introduce bottleneck features, which act as independent features bridging the gaps between the features of the two modalities \cite{nagrani_2021_bottleneck}. The reference modality is lower utilized compared to TransFusion as it does not allow complete mutual communication of the two modalities. Mathematically, let $\bd{B} \in \R^{n_b\times d}$ denote the bottleneck features, the \textbf{B}ottleneck \textbf{T}rans\textbf{F}usion (BTF) expert is given as,
\begin{equation}
    \operatorname{BTF}(\bd{F}_1, \bd{F}_2) = \operatorname{SA}_1(\bd{F}_1, \operatorname{SA}_2(\bd{B}, \bd{F}_2)[:n_b,:])[:n_1,:].
\end{equation}

\paragraph{SNNFusion.} The \textbf{S}NN\textbf{F}usion (SF) expert is designed for fusion that is genomic dominant, as it has promising results when applied solely to genomic data. The utilization of reference modality is even lower when adopting this expert. In SNNFusion, there are two SNNs \cite{klambauer_self_2017} and our SF expert is given as,
\begin{equation}
    \operatorname{SF}(\bd{F}_1, \bd{F}_2) = \operatorname{SNN}_1(\operatorname{\operatorname{RMSNorm}(\bd{F}_1)}) + \operatorname{mean}(\operatorname{SNN}_2(\operatorname{\operatorname{RMSNorm}(\bd{F}_2)})),
\end{equation}
where $\operatorname{SNN}(\cdot)$ consists of a linear layer, an \textbf{E}xponential \textbf{L}inear \textbf{U}nit (ELU) activation layer \cite{clevert2016fast} and an alpha dropout layer \cite{klambauer_self_2017}. 

\paragraph{DropF2Fusion.} This expert drops $\bd{F}_2$ during fusion, serving as a skip layer, as it completely stops utilizing $\bd{F}_2$. It is particularly useful when using one modality is accurate enough. Mathematically, the \textbf{D}ropF2\textbf{F}usion (DF) expert is given as,

\begin{equation}
    \operatorname{DF}(\bd{F}_1, \bd{F}_2) = \bd{F}_1.
\end{equation}

\subsection{Why Self-attention over Co-attention}\label{sec:transformer}

Mathematically, the \textbf{C}o-\textbf{A}ttention (CA) and SA can be given as,
\begin{equation}\label{eq:coattn}
    \operatorname{CA}(\bd{F}_1, \bd{F}_2) = \operatorname{Softmax}(\frac{(\bd{F}_1\bd{Q})(\bd{F}_2\bd{K})^T}{\sqrt{d}})(\bd{F}_2\bd{V}).
\end{equation}
\begin{equation}\label{eq:selfattn}
    \begin{aligned}
    \operatorname{SA}([\bd{F}_1, \bd{F}_2]) & = \operatorname{Softmax}(\frac{\begin{bmatrix}
    \bd{F}_1\bd{Q} \\
    \bd{F}_2\bd{Q} \\
\end{bmatrix}\begin{bmatrix}
    (\bd{F}_1\bd{K})^T & (\bd{F}_2\bd{K})^T\\
\end{bmatrix}}{\sqrt{d}})\begin{bmatrix}
    \bd{F}_1\bd{V} \\
    \bd{F}_2\bd{V} \\
\end{bmatrix},\\
 & = \operatorname{Softmax}(\frac{\begin{bmatrix}
    (\bd{F}_1\bd{Q})(\bd{F}_1\bd{K})^T & \textcolor{red}{(\bd{F}_1\bd{Q})(\bd{F}_2\bd{K})^T} \\
    (\bd{F}_2\bd{Q})(\bd{F}_1\bd{K})^T & (\bd{F}_2\bd{Q})(\bd{F}_2\bd{K})^T \\
\end{bmatrix}}{\sqrt{d}})\begin{bmatrix}
    \bd{F}_1\bd{V} \\
    \textcolor{red}{\bd{F}_2\bd{V}} \\
\end{bmatrix}.\\
\end{aligned}
\end{equation}

The matrix multiplication result of the red font portion in \cref{eq:selfattn} matches $\operatorname{CA}(\bd{F}_1, \bd{F}_2)$ in \cref{eq:coattn}. $\operatorname{CA}(\bd{F}_2, \bd{F}_1)$, $\operatorname{SA}(\bd{F}_1)$ and $\operatorname{SA}(\bd{F}_2)$ are also embedded in \cref{eq:selfattn}. Hence, we can conclude that CA is a sub-optimal substitute for SA. Therefore, we design the experts based on self-attention instead of co-attention.

\section{Experiments and Results}
\begin{table}[t]
\begin{center}
\caption{C-index Results for different methods on three TCGA datasets. The best results are underlined in red, while the second best are italicized in blue. ``Geno.'' denotes the utilization of genomic profiles, and ``Patho.'' signifies the use of WSIs.}
\begin{tabular}{l|cc|cccc}
\hline
\multirow{2}{*}{} & \multicolumn{2}{c|}{\bf Modality} & \multicolumn{4}{c}{\bf Dataset}\\
& \textbf{Geno.} &  \textbf{Patho.} & \textbf{BLCA} & \textbf{UCEC} & \textbf{LUAD} & \textbf{Overall}\\
        \hline \hline
        SNN \cite{klambauer_self_2017} & \boldcheckmark & & 0.618$\pm$0.022 & 0.679$\pm$0.040 & 0.611$\pm$0.047& 0.636\\
        SNNTrans \cite{klambauer_self_2017} & \boldcheckmark & & 0.659$\pm$0.032 & 0.656$\pm$0.038 & 0.638$\pm$0.022 & 0.651\\
        \hline
        AttnMIL \cite{ilse_attention-based_2018} & & \boldcheckmark & 0.599$\pm$0.048 & 0.658$\pm$0.036 & 0.620$\pm$0.061 & 0.626\\
        CLAM-SB \cite{lu_data-efficient_2021} & & \boldcheckmark & 0.559$\pm$0.034 & 0.644$\pm$0.061 & 0.594$\pm$0.063 & 0.599\\
        CLAM-MB \cite{lu_data-efficient_2021} & & \boldcheckmark & 0.565$\pm$0.027 & 0.609$\pm$0.082 & 0.582$\pm$0.072  & 0.585\\
        TransMIL \cite{shao_transmil} & & \boldcheckmark & 0.575$\pm$0.034 & 0.655$\pm$0.046 & 0.642$\pm$0.046 & 0.624\\
        DTFD-MIL \cite{zhang_dtfd-mil_2022} & & \boldcheckmark & 0.546$\pm$0.021 & 0.656$\pm$0.045 & 0.585$\pm$0.066 & 0.596\\
        \hline
        MCAT \cite{chen2021multimodal} & \boldcheckmark & \boldcheckmark & 0.672$\pm$0.032 & 0.649$\pm$0.043 & 0.659$\pm$0.027 & 0.660\\
        Porpoise \cite{chen_2022_pancancer} & \boldcheckmark & \boldcheckmark & 0.636$\pm$0.024 & \sbest{0.692}$\pm$0.048 & 0.647$\pm$0.031 & 0.658\\
        MOTCAT \cite{xu_2023_motcat} & \boldcheckmark & \boldcheckmark & \sbest{0.682}$\pm$0.023 & 0.671$\pm$0.053 & 0.673$\pm$0.040 & 0.675\\
        CMTA \cite{zhou2023cross} & \boldcheckmark & \boldcheckmark & 0.672$\pm$0.038 & 0.691$\pm$0.066 & \sbest{0.676}$\pm$0.037 & \sbest{0.680}\\
        \hline \hline
        \rowcolor{LightCyan}
        MoME (Ours) & \boldcheckmark & \boldcheckmark & \best{0.686}$\pm$0.041 & \best{0.706}$\pm$0.038 & \best{0.691}$\pm$0.040 & \best{0.694}\\
        \hline
\end{tabular}
\label{tab:main_result}
\end{center}
\end{table}
\begin{table}[t]
\begin{center}
\caption{Ablation study results for using different experts in our MoME (left) and sensitivity analysis in the number of bottleneck features (right). ``T.'' represents TransFusion, ``B.'' represents Bottleneck TransFusion, ``S.'' represents SNNFusion, ``D.'' represents DropF2Fusion, and ``\#B.'' denotes the number of bottleneck features.}
\begin{tabular}{l|cccc|cc||c|cc}
\hline
\multirow{2}{*}{} & \multicolumn{4}{c|}{\bf Experts} & \multicolumn{2}{c||}{\bf Dataset} & \multirow{2}{*}{\bf \#B.} & \multicolumn{2}{c}{\bf Dataset}\\
& \textbf{T.} &  \textbf{B.} & \textbf{S.} & \textbf{D.} & \textbf{UCEC} & \textbf{LUAD} & & \textbf{UCEC} & \textbf{LUAD}\\
        \hline \hline
        MoME & \boldcheckmark & \boldcheckmark & \boldcheckmark & \boldcheckmark& \sbest{0.706}$\pm$0.038 & \best{0.691}$\pm$0.040 & 1 & 0.690$\pm$0.029 & 0.669$\pm$0.029 \\
        MoME &  & \boldcheckmark & \boldcheckmark & \boldcheckmark & 0.693$\pm$0.041 & 0.673$\pm$0.043 & 2 & \sbest{0.706}$\pm$0.038 & \best{0.691}$\pm$0.040 \\
        MoME & \boldcheckmark &  & \boldcheckmark & \boldcheckmark & 0.703$\pm$0.052 & 0.663$\pm$0.064 & 4 & 0.699$\pm$0.053 & \sbest{0.673}$\pm$0.040 \\
        MoME & \boldcheckmark & \boldcheckmark & & \boldcheckmark & \best{0.738}$\pm$0.060 & 0.669$\pm$0.045 & 8 & \best{0.717}$\pm$0.072 & 0.658$\pm$0.035 \\
        MoME & \boldcheckmark & \boldcheckmark & \boldcheckmark & & 0.684$\pm$0.053 & \sbest{0.685}$\pm$0.047 & 16 & 0.704$\pm$0.044 & 0.662$\pm$0.034 \\
        \hline
        TF & \boldcheckmark & &  & & 0.690$\pm$0.059 & 0.655$\pm$0.061 & / & / & /\\
        \hline
\end{tabular}
\label{tab:ablation}
\end{center}
\end{table}
\subsection{Datasets}
\textbf{T}he \textbf{C}ancer \textbf{G}enome \textbf{A}tlas (TCGA\footnote{https://www.cancer.gov/ccg/research/genome-sequencing/tcga}) project provides extensive information on patients under study, including WSIs, genomic data, and ground truth survival time. The datasets used in our experiments include 373 samples of \textbf{B}ladder \textbf{U}rothelial \textbf{CA}rcinoma (BLCA), 480 samples of \textbf{U}terine \textbf{C}orpus \textbf{E}ndometrial \textbf{C}arcinoma (UCEC), and 453 samples of \textbf{LU}ng \textbf{AD}enocarcinoma (LUAD).
\subsection{Implementation Details}

\subsubsection{Training Settings.} We select a wide range of baseline methods, including those focused on genomic data, WSIs, and both modalities. The methods for comparison are: SNN~\cite{klambauer_self_2017}, SNNTrans~\cite{klambauer_self_2017}, AttnMIL~\cite{ilse_attention-based_2018}, CLAM~\cite{lu_data-efficient_2021}, TransMIL~\cite{shao_transmil}, DTFD-MIL~\cite{zhang_dtfd-mil_2022}, Porpoise~\cite{chen_2022_pancancer},  MCAT~\cite{chen2021multimodal}, MOTCAT~\cite{xu_2023_motcat} and CMTA~\cite{zhou2023cross}. We choose the \textbf{C}oncordance index (C-index), a commonly employed metric in survival analysis, as our evaluation metric. To evaluate the performance of these methods, we conduct a five-fold cross-validation. Each model is training for 20 epochs, and the best validation performance obtained among these epochs is considered as the final performance for the respective fold. The means and standard deviations of the C-index for each method on different datasets are reported.

\subsubsection{Hyper-parameters.} Adam \cite{kingma_adam_2015} optimizer is used in our experiment. The learning rate and weight decay are set to 2$\times$10$^{-4}$ and 1$\times$10$^{-5}$, respectively \cite{xu_2023_motcat}. The WSIs are split into patches sized of 256 $\times$ 256 pixels at 20$\times$ magnification and ResNet-50 \cite{resnet_he_16} pre-trained on ImageNet is used to extract features from them. The number of bottleneck features in our experiments $n_b$ is 2. We utilize the micro-batch technique \cite{xu_2023_motcat} and the size of the micro-batch is 4,096.

\subsection{Comparison Results}
We conduct a comprehensive comparison of our method with both unimodal methods and other state-of-the-art multimodal methods. The results are presented in \cref{tab:main_result}. Our method consistently outperforms all other methods across all datasets, particularly on the UCEC and LUAD datasets, where it exhibits a significant performance advantage over previous approaches. Our method achieves improvements of 0.4\%, 1.4\%, and 1.5\% on the C-index of the three datasets, respectively, compared to previous methods, as well as a 1.4\% improvement in overall performance. These results suggest that our MoME is applicable to general survival analysis settings. Notably, despite their simple structures, methods based on genomic data outperform those based on WSIs, highlighting the importance of genomic data in survival analysis. Furthermore, multimodal methods consistently outperform unimodal methods, further demonstrating the efficacy of multimodal approaches and the necessity of incorporating multiple modalities.

\subsection{Ablation Studies}

\subsubsection{Choices of Experts.} We conduct experiments to assess the effectiveness of each expert by deactivating them individually. The experimental results are presented in the left part of \cref{tab:ablation}. We observe that our MoME with all experts achieves the best performance on LUAD and the second-best performance on UCEC. The MoME without the SNNFusion expert achieves the best performance on UCEC. These results indicate that the most crucial components for UCEC and LUAD differ, suggesting that a MoME focusing on a specific modality could be beneficial. This further supports the necessity of our MoME, which dynamically selects different experts for different samples. Additionally, the performance of our MoME compared to TransFusion reaffirms its superiority.

\subsubsection{Sensitivity Analysis on BTF.} We conduct a sensitivity analysis on UCEC and LUAD by varying the number of bottleneck features used in BTF from 1 to 16. The results are presented in the right part of \cref{tab:ablation}. We observe that our MoME achieves the overall best performance when the number of bottleneck features is 2, which strikes a balance between these two datasets. Specifically, our MoME achieves the best performance on UCEC when the number of bottleneck features is 8, however, its performance on LUAD is the lowest.

\section{Conclusion}
In this paper, we introduced a BPE paradigm and a MoME layer for cancer survival analysis. The BPE paradigm enables deep fusion by performing feature encoding and fusion simultaneously, leveraging one modality as a reference to encode the other. With this, our BPE can addresses the severe heterogeneity between WSI and genomic features. Additionally, our MoME layer dynamically selects the most appropriate expert to model the intricate inter- and intra-modal correlations, addressing the challenges posed by variations in key features. Through extensive experiments, we demonstrated that our method outperforms other multimodal learning approaches in survival prediction, and the results suggest that our method could be applied to a general survival analysis setting. 

\subsubsection{\ackname} The research presented in this paper was partially supported by the Research Grants Council of the Hong Kong Special Administrative Region, China (CUHK 14222922, RGC GRF 2151185).
\subsubsection{\discintname}
The authors have no competing interests to declare that are relevant to the content of this article.
\newpage
%
%
%
\bibliographystyle{splncs04}
\bibliography{main}
\end{document}